\documentclass[conference]{IEEEtran}
\IEEEoverridecommandlockouts

\usepackage{amsmath}
\usepackage{amssymb,pifont,wasysym,graphicx,booktabs,makecell}
\usepackage{amsmath,float,stfloats,multirow,array}
\usepackage{tikz,pgfplots}
\usepackage[hidelinks]{hyperref}
\pgfplotsset{compat=1.18}
\usetikzlibrary{shapes.geometric,arrows.meta,positioning,
                fit,backgrounds,calc,patterns}

\usepackage[utf8]{inputenc}
\usepackage[T1]{fontenc}

\usepackage[
  style=numeric-comp,
  natbib=true,
  maxbibnames=2,
  minbibnames=1,
  maxcitenames=1,
  mincitenames=1,
  doi=false,
  url=false,
  isbn=false,
  eprint=false,
  giveninits=true,
]{biblatex}
\AtEveryBibitem{\footnotesize\clearfield{note}\clearfield{series}\clearfield{number}}
\DeclareFieldFormat{labelnumberwidth}{\footnotesize[#1]}

\usepackage{caption}
\title{Privacy-Preserving Federated Autoencoder for ECG
       Anomaly Detection on Edge Devices}

\author{
\IEEEauthorblockN{
\begin{tabular}{c}
Kaan Arda Akyol, Jakub Kacper Szel\k{a}g, Aydin Abadi, Maha Alghamdi, Ghadah Albalawi,\\
Ghouse Ibrahim Kaleelullah, Hilal Tutus, Sarah Al Subaiei, Shardul Kapse,\\
Syed Mohammed Raheeb, Mujeeb Ahmed, Rehmat Ullah
\end{tabular}
}
\vspace{2pt}
\IEEEauthorblockA{
School of Computing, Newcastle University, Newcastle upon Tyne, United Kingdom\\
Email: \{K.A.Akyol2, J.K.Szelag2, Aydin.Abadi, M.Alghamdi2, G.S.S.Albalawi2,\\
G.M.Ibrahim-Kaleelullah2, H.Tutus2, S.S.A.Alsubaiei2, S.Kapse2,\\
S.M.Mohammed-Raheeb2, Mujeeb.Ahmed, Rehmat.Ullah\}@newcastle.ac.uk
}
}

\begin{document}
\maketitle

\begin{abstract}
Continuous electrocardiography (ECG) monitoring could surface rhythm abnormalities before they escalate into cardiovascular events. However, a deployable system must satisfy three requirements simultaneously: legal-grade privacy (GDPR, HIPAA), real-time inference on constrained edge hardware, and detection quality under non-IID cross-hospital data. 

We design and evaluate an end-to-end federated system addressing all three for unsupervised 12-lead ECG anomaly detection on PTB-XL dataset, combining three autoencoder families (VanillaAE, ConvAE, VAE), Flower-based federated averaging (FedAvg) across ten simulated hospitals, client-side differentially private SGD (DP-SGD) with a R\'enyi-DP accountant, and 8-bit integer (INT8) post-training quantization with Raspberry~Pi~4 benchmarking. Our main contributions are: an empirical characterization of how these mechanisms compose, practical DP-specific recommendations, as well as technical and security insights to an otherwise clinically sensitive subject, all while prioritizing ethical considerations. Federated learning matches or exceeds the centralized baseline across all architectures (ConvAE federated area under the ROC curve, AUROC, $0.782$), and an $\varepsilon$ sweep identifies $\varepsilon{=}4$ as the recommended clinical operating point. INT8 quantization roughly halves model size and cuts Pi~4 latency by up to $44\%$ with ${<}0.12\%$ AUROC loss. Crucially, DP and quantization penalties are empirically independent, so practitioners need not trade a strong privacy guarantee for a compact edge footprint. To our knowledge, this is the first system combining federated learning, formal $(\varepsilon,\delta)$-DP, unsupervised reconstruction-based detection, and quantized AArch64 deployment.
\end{abstract}

\renewcommand\IEEEkeywordsname{Keywords}

\vspace{-2mm}
\section{Introduction}
\label{sec:intro}

Cardiovascular disease is the world's leading cause of mortality, responsible for an estimated 17.9 million deaths annually~\cite{WHO_CVD}, and a substantial fraction are preceded by detectable rhythm abnormalities that continuous electrocardiography (ECG) surveillance could catch before escalation. Wearable and implantable monitors promise real-time cardiac screening that bridges onset and intervention~\cite{Rah_Ref11}, shifting cardiology from reactive treatment toward proactive prevention. Turning that promise into a deployable system, however, requires three properties simultaneously.

\emph{Privacy compliance.} Patient ECG data is among the most sensitive categories under GDPR (EU~2016/679) and HIPAA, and cannot be freely centralized. Federated learning (FL)~\cite{Exp_Set6} keeps raw records local and shares only model updates, and has been increasingly adopted in distributed medical systems~\cite{Rah_Ref11,Kaissis2020}. FL alone is insufficient: gradient inversion, membership inference, and de-anonymization attacks can reconstruct training records or re-identify individuals from shared updates or deployed ECG models~\cite{Paper5,ECG_Attacks2025,Deanon_ECG2025}. Differential privacy (DP) via differentially private stochastic gradient descent (DP-SGD)~\cite{Abadi2016,Exp_Set4} closes this gap with a rigorous $(\varepsilon,\delta)$-guarantee~\cite{DP_Survey2024}, but its utility cost on reconstruction-based anomaly detectors, which rely on subtle mean-squared-error (MSE) differences between normal and pathological waveforms, remains poorly characterized.

\emph{Edge efficiency.} Continuous monitoring devices are overwhelmingly AArch64-based, typically shipping with ARM Cortex-A cores~\cite{Paper13,Paper2}, and the Raspberry~Pi~4 is the canonical research proxy for this class. 8-bit integer (INT8) post-training quantization (PTQ) can halve model size with negligible accuracy loss in supervised settings~\cite{Comput_Result}, yet its effect on the fine-grained reconstruction fidelity that drives ECG anomaly scoring has not been validated on AArch64 hardware.

\emph{Detection effectiveness.} Any model satisfying the first two requirements must still maintain sufficient detection quality under non-independent and identically distributed (non-IID) hospital-scale partitioning for population-level screening. Satisfying each property in isolation is tractable; satisfying all three end-to-end on real edge hardware is where the literature is currently sparse.

Simply stacking FedAvg, DP-SGD, and INT8 PTQ is insufficient for two reasons. First, DP noise interacts with reconstruction error differently than with classification cross-entropy: supervised pipelines tolerate moderate DP-SGD noise because cross-entropy provides strong per-class gradients, whereas reconstruction-based detectors exhibit a sharp utility floor below which noise overwhelms the signal. Second, PTQ has not been validated on the reconstruction fidelity required for ECG anomaly scoring on AArch64. These gaps motivate a fully evaluated end-to-end system.

The proposed framework targets two complementary dimensions within a unified federated pipeline: (i)~security and privacy, through formal $(\varepsilon,\delta)$-DP at the client level, and (ii)~computational efficiency, through INT8 PTQ and on-device benchmarking on AArch64 edge hardware.

\vspace{-2mm}
\subsection{Our Contributions}

We design, implement, and evaluate an end-to-end federated system addressing privacy compliance, edge efficiency, and detection effectiveness in a single unified pipeline. Our contributions are as follows. 
\begin{enumerate}\setlength{\itemsep}{1pt}\setlength{\parskip}{0pt}

\item \textbf{Federation effect.} Federated training matches the centralized baseline across three autoencoder families under non-IID partitioning ($K{=}10$, Dirichlet $\alpha{=}0.5$). ConvAE gains $+0.018$ area under the receiver operating characteristic curve (AUROC) while VAE gains $+0.165$, with the larger VAE gain attributable to federation-induced regularization of a collapsed baseline rather than an intrinsic architectural advantage.

\item \textbf{Privacy-utility trade-off.} A five-point $\varepsilon$ sweep identifies $\varepsilon{=}4$ as a principled operating point for clinical data and empirically locates a utility floor at AUROC${\approx}0.610$ below which gradient noise overwhelms reconstruction learning.

\item \textbf{Edge deployment.} INT8~PTQ reduces model size by ${\approx}53\%$ and Raspberry~Pi~4 latency by $16.6$--$44.3\%$ (a $1.20$--$1.80\times$ speedup) with ${<}\,0.12\%$ AUROC degradation across all three architectures.

\item \textbf{Combined configuration.} FL\,+\,DP\,+\,INT8 is, to our knowledge, the first evaluated configuration of its kind for unsupervised ECG. DP and quantization penalties are empirically independent, so enabling both simultaneously is strictly preferable for edge deployment.
\end{enumerate}
As Table~\ref{tab:literature_compare} shows, no prior work addresses all nine evaluated dimensions simultaneously; to our knowledge this is the first evaluated configuration of the four-way intersection (FL + formal DP + unsupervised reconstruction-based detection + quantized AArch64 deployment) on ECG data.

We also report a secondary finding on data normalization: switching from per-lead min-max scaling to $z$-score normalization raised centralized ConvAE AUROC from ${\approx}0.55$ to $0.795$. Normalization choices otherwise harmless in supervised settings can be destructive for anomaly scoring, and we adopt $z$-score normalization throughout.


\section{Related Work}
\label{sec:related}

Building a system that satisfies privacy, efficiency, and effectiveness together requires synthesizing four otherwise disjoint threads: unsupervised ECG anomaly detection, federated learning for ECG, privacy-preserving FL with edge deployment, and differential privacy in healthcare FL.

\textbf{Unsupervised ECG anomaly detection.} Autoencoder-based architectures dominate centralized evaluation via reconstruction error as the anomaly signal. A community-aware unsupervised pipeline~\cite{Paper1} achieves AUC\,$=$\,$0.836$ but operates on generic time-series rather than 12-lead ECG. A transformer-based model~\cite{Paper8} reports strong accuracy on MIT-BIH and ECG5000 but remains centralized and computationally heavy. \citet{Paper13} target edge deployment via TinyML with pruning and PTQ, but consider neither federated training nor formal DP. A comparative VAE-BiLSTM study~\cite{Paper11} is limited to single-dataset centralized evaluation, and a recent survey~\cite{Paper14} confirms that autoencoders and VAEs remain dominant label-free architectures while highlighting the need for larger, distributed training corpora.

\textbf{Federated learning for ECG analysis.} A 1D ResNet-34 trained with FedOpt~\cite{Paper3} across multi-institutional PhysioNet partitions achieves F1\,$=$\,$0.93$, and a federated anomaly detector with global threshold derivation~\cite{Paper15} shows improvement under non-IID conditions, yet both omit DP and edge evaluation. \citet{Paper6} report a federated macro-F1 of $0.58$ against a local-training $0.63$, quantifying the accuracy cost of federation alone. \citet{Paper7} explore unsupervised FL anomaly detection on brain MRI rather than ECG, so its findings do not directly transfer.

\textbf{Privacy-preserving FL with edge deployment.} This thread combines federated training with compression for on-device inference, typically without formal DP. A quantized DNN on edge hardware~\cite{Paper2} demonstrates feasibility without formal privacy; Gramian Angular Field representations with FL on heterogeneous IoT devices~\cite{Paper12} address device heterogeneity but not privacy; a single-lead federated framework~\cite{Paper9} achieves sensitivity\,$=$\,$0.88$ and specificity\,$=$\,$0.84$ without DP; and a federated denoising autoencoder with Grad-CAM~\cite{Paper4} reaches $94.5\%$ accuracy on Raspberry~Pi but omits DP and energy metrics. \citet{XAI_FL_Edge2025} combine FL, compression, and explainability on edge hardware but do not provide formal privacy guarantees.

\textbf{Differential privacy in healthcare FL.} DP has become a de-facto standard for ML in healthcare, but its integration with ECG analysis remains shallow. \citet{Abadi2016} established DP-SGD as the canonical per-example mechanism; \citet{Mironov2017} tightened its accounting via R\'enyi-DP; \citet{McMahan2018} extended it to federated training; and \citet{DP_Survey2024} survey the modern landscape. In clinical imaging, \citet{Kaissis2020} argue that secure, private FL is a prerequisite for institutional adoption, and \citet{Ponomareva2023} provide deployment heuristics for $\varepsilon$ selection. Within ECG specifically, \citet{Paper5} train a supervised 1D ResNet on 1.56M recordings under DP-SGD, reporting AUROC${\approx}0.81$; \citet{Zhang2024JBHI} propose a two-stage end-edge-cloud DP scheme; \citet{Bokhari2025} fuse personalized FL with local DP on 12-lead ECG; and \citet{DP_XAI_ECG2025} pair DP with explainability for 12-lead classification. Crucially, every cited ECG+DP work is \emph{supervised}: none applies DP to reconstruction-based unsupervised detection, where the weaker per-sample gradient signal interacts with injected noise in qualitatively different ways.

\textbf{Gap at the intersection.} Taken together, the literature shows mature coverage of each pair of threads; however, the four-way intersection required for a practically deployable system (comprising FL for distributed healthcare, formal client-level DP, unsupervised reconstruction-based detection, and quantized AArch64 deployment) remains unaddressed. Table~\ref{tab:literature_compare} consolidates this analysis across nine evaluation dimensions.

\begin{table*}[t]
\centering
\caption{Comparison of literature features in federated ECG studies. \checkmark\,=\,explicitly evaluated, \ding{55}\,=\,not considered, \LEFTcircle\,=\,partial. FL: Federated Learning; DP: Differential Privacy; AD: Anomaly Detection.}
\label{tab:literature_compare}
\setlength{\tabcolsep}{8pt}
\normalfont\footnotesize
\begin{tabular}{lcccccccccc}
\toprule
\textbf{Paper} &
\textbf{FL} &
\textbf{DP} &
\makecell{\textbf{Unsup}\\\textbf{AD}} &
\makecell{\textbf{Edge}\\\textbf{Deploy}} &
\makecell{\textbf{ECG}\\\textbf{Data}} &
\makecell{\textbf{12}\\\textbf{Lead}} &
\makecell{\textbf{Model}\\\textbf{Comp.}} &
\makecell{\textbf{Non-}\\\textbf{IID}} &
\textbf{Energy}\\
\midrule
\citet{Paper1}  & \checkmark & \ding{55} & \checkmark & \ding{55} & \ding{55} & \ding{55} & \ding{55} & \checkmark & \ding{55} \\
\citet{Paper2}  & \ding{55} & \ding{55} & \ding{55} & \checkmark & \checkmark & \ding{55} & \ding{55} & \ding{55} & \checkmark \\
\citet{Paper3}  & \checkmark & \ding{55} & \ding{55} & \ding{55} & \checkmark & \checkmark & \checkmark & \LEFTcircle & \ding{55} \\
\citet{Paper4}  & \LEFTcircle & \ding{55} & \ding{55} & \checkmark & \checkmark & \ding{55} & \ding{55} & \ding{55} & \LEFTcircle \\
\citet{Paper5}  & \checkmark & \checkmark & \ding{55} & \ding{55} & \checkmark & \checkmark & \ding{55} & \ding{55} & \ding{55} \\
\citet{Paper6}  & \checkmark & \ding{55} & \ding{55} & \ding{55} & \checkmark & \checkmark & \checkmark & \checkmark & \ding{55} \\
\citet{Paper7}  & \checkmark & \ding{55} & \checkmark & \ding{55} & \ding{55} & \ding{55} & \ding{55} & \checkmark & \ding{55} \\
\citet{Paper8}  & \LEFTcircle & \ding{55} & \checkmark & \ding{55} & \checkmark & \ding{55} & \checkmark & \ding{55} & \ding{55} \\
\citet{Paper9}  & \checkmark & \ding{55} & \checkmark & \checkmark & \checkmark & \ding{55} & \ding{55} & \checkmark & \ding{55} \\
\citet{Paper10} & \checkmark & \ding{55} & \ding{55} & \ding{55} & \checkmark & \ding{55} & \ding{55} & \checkmark & \ding{55} \\
\citet{Paper11} & \ding{55} & \ding{55} & \checkmark & \ding{55} & \checkmark & \checkmark & \checkmark & \ding{55} & \ding{55} \\
\citet{Paper12} & \checkmark & \ding{55} & \ding{55} & \checkmark & \checkmark & \ding{55} & \ding{55} & \checkmark & \checkmark \\
\citet{Paper13} & \ding{55} & \ding{55} & \checkmark & \checkmark & \checkmark & \ding{55} & \checkmark & \ding{55} & \checkmark \\
\citet{Paper14} & \ding{55} & \ding{55} & \checkmark & \ding{55} & \checkmark & \LEFTcircle & \ding{55} & \ding{55} & \ding{55} \\
\citet{Paper15} & \checkmark & \ding{55} & \checkmark & \ding{55} & \ding{55} & \ding{55} & \ding{55} & \checkmark & \ding{55} \\
\textbf{Our Approach} & \checkmark & \checkmark & \checkmark & \checkmark & \checkmark & \checkmark & \checkmark & \checkmark & \LEFTcircle \\
\bottomrule
\end{tabular}
\end{table*}

\section{Preliminaries}
\label{sec:prelim}

This section introduces the four core building blocks used throughout the paper: federated learning, differential privacy, DP-SGD, and the diagnostic vocabulary of 12-lead ECG, pitched at a level readable across healthcare, confidential computing, and AI.

\textbf{Federated Learning (FL) and FedAvg.} FL is a distributed training paradigm in which $K$ clients (e.g., hospitals) train a shared model under a central server without exchanging raw data~\cite{Exp_Set6}. Each client holds local dataset $\mathcal{D}_k$ of size $n_k$ with $N{=}\sum_k n_k$, minimizing $F(w){=}\sum_{k=1}^{K}(n_k/N) F_k(w)$, where the local objective $F_k(w){=}\mathbb{E}_{\xi\sim\mathcal{D}_k}[\ell(w;\xi)]$ is the expected per-sample loss, with $\mathbb{E}[\cdot]$ as the expectation, $\ell$ the reconstruction loss, and $\xi$ a training example drawn from $\mathcal{D}_k$. FedAvg~\cite{Exp_Set6} is the canonical algorithm: at round $t$, selected clients perform $E$ local SGD epochs to obtain $w_k^{t+1}$, and the server aggregates via:
\begin{equation}
    w_{t+1} = \sum_{k=1}^{K} \frac{n_k}{N}\, w_k^{t+1} .
    \label{eq:fedavg}
\end{equation}

\textbf{$(\varepsilon,\delta)$-Differential Privacy.} A randomized mechanism $\mathcal{M}$ satisfies $(\varepsilon,\delta)$-DP if, for any two datasets $D,D'$ differing in a single record and any measurable $S{\subseteq}\text{Range}(\mathcal{M})$:
\begin{equation}
  \Pr[\mathcal{M}(D)\!\in\!S] \;\leq\; e^{\varepsilon}\,\Pr[\mathcal{M}(D')\!\in\!S] + \delta.
  \label{eq:dp}
\end{equation}
Smaller $\varepsilon$ yields stronger privacy; $\delta$ bounds the probability of catastrophic leakage and is conventionally set to ${<}\,1/N$~\cite{DworkRoth2014}. This work uses the client-side (local DP) variant, in which each client randomizes its update before release, rather than central DP where the server adds noise to aggregates.

\textbf{DP-SGD and DP-FedAvg.} DP-SGD~\cite{Abadi2016,Exp_Set4} clips per-sample gradients $g_i{=}\nabla\ell(w;\xi_i)$ to $\ell_2$-norm $C$ and injects calibrated Gaussian noise:
\begin{equation}
  \tilde{g} \;=\; \tfrac{1}{B}\!\left(\sum_{i\in\mathcal{B}}
    g_i\!\cdot\!\min\!\left(1,\tfrac{C}{\|g_i\|_2}\right)
    + \mathcal{N}(0,\sigma^2 C^2 \mathbf{I})\right),
  \label{eq:dpsgd}
\end{equation}
where $\mathcal{B}$ is the minibatch of size $B$, $\sigma$ the noise multiplier, and $\mathbf{I}$ the identity matrix. The R\'enyi-DP accountant~\cite{Mironov2017} composes per-step R\'enyi divergences across all $R{\cdot}E$ steps, yielding tighter bounds than the strong composition theorem. DP is applied to per-sample gradients at each client, protecting the learned parameters while leaving local data untouched. In DP-FedAvg~\cite{McMahan2018}, clients run Eq.~\eqref{eq:dpsgd} locally and the server aggregates the noisy updates. The privacy unit is a single ECG recording (example-level DP per client, composed over $R{=}50$ rounds).

\textbf{ECG diagnostic categories.} PTB-XL annotates recordings with SCP-ECG codes grouped into five superclasses: \emph{NORM} (normal sinus rhythm) and four abnormal classes, \emph{MI} (myocardial infarction), \emph{STTC} (ST/T-wave changes), \emph{HYP} (hypertrophy), and \emph{CD} (conduction disturbances). Models are trained on NORM samples only and flag any recording whose reconstruction error exceeds a threshold as anomalous; the abnormal classes are never seen during training.

\section{Methodology}
\label{sec:methodology}

The pipeline takes raw 12-lead ECG recordings as input and produces a deployed INT8 anomaly detector on an edge device, passing through four stages: data preparation, federated learning, differential privacy integration, and post-training deployment. Design rationale is deferred to Section~\ref{sec:discussion}.

\vspace{-1mm}
\subsection{System Architecture}
\label{sec:arch}

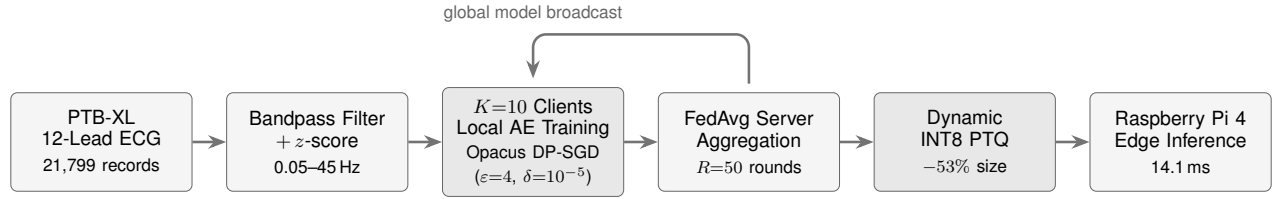
\begin{figure*}[!t]
\centering
\resizebox{0.92\textwidth}{!}{%
\begin{tikzpicture}[
  font=\sffamily\small,
  node distance=0.55cm,
  box/.style={draw=black!60, rounded corners=3pt, fill=#1,
              align=center, minimum height=1.6cm,
              text width=2.6cm, inner sep=5pt},
  box/.default={gray!8},
  arr/.style={-Stealth, very thick, draw=black!55},
  lbl/.style={font=\sffamily\footnotesize, text=black!60},
]
  \node[box]                 (ptb)  {PTB-XL\\12-Lead ECG\\[2pt]\footnotesize 21{,}799 records};
  \node[box, right=of ptb]          (pre)  {Bandpass Filter\\$+$\,$z$-score\\[2pt]\footnotesize 0.05--45\,Hz};
  \node[box=gray!18, right=of pre]  (cl)   {$K{=}10$ Clients\\Local AE Training\\[2pt]\footnotesize Opacus DP-SGD\\[1pt]\footnotesize ($\varepsilon{=}4$, $\delta{=}10^{-5}$)};
  \node[box, right=of cl] (srv)  {FedAvg Server\\Aggregation\\[2pt]\footnotesize $R{=}50$ rounds};
  \node[box=gray!18, right=of srv]  (ptq)  {Dynamic\\INT8 PTQ\\[2pt]\footnotesize $-53\%$ size};
  \node[box, right=of ptq]          (rpi)  {Raspberry Pi 4\\Edge Inference\\[2pt]\footnotesize 14.1\,ms};

  \draw[arr] (ptb) -- (pre);
  \draw[arr] (pre) -- (cl);
  \draw[arr] (cl)  -- (srv);
  \draw[arr] (srv) -- (ptq);
  \draw[arr] (ptq) -- (rpi);

  \draw[arr, rounded corners=6pt]
    ([yshift=4pt]srv.north) -- ++(0,0.8) -| ([yshift=4pt]cl.north)
    node[lbl, midway, above, yshift=2pt] {global model broadcast};
\end{tikzpicture}}
\caption{Proposed pipeline. PTB-XL ECGs are filtered, normalized, and split across 10 non-IID federated clients. Local autoencoders, with optional DP-SGD via Opacus, are trained and aggregated by Flower FedAvg over 50 rounds. The best 32 bit floating point (FP32) model is quantized to INT8 and benchmarked on a Raspberry Pi 4 (14.1 ms).}
\label{fig:AR}
\end{figure*}

The pipeline shown in Fig.~\ref{fig:AR} comprises four stages.

\textit{Data preparation.} PTB-XL recordings are bandpass-filtered ($0.05$ to $45$\,Hz, fourth-order zero-phase Butterworth) and per-lead $z$-score normalized using training-set statistics only ($\hat{x}_l = (x_l - \mu_l)/\sigma_l$). A patient-level 70/15/15 split (seed\,$42$) yields $6{,}294$ normal training samples distributed across $K{=}10$ clients via a Dirichlet draw ($\alpha{=}0.5$), producing a $75.7\times$ volume disparity between the largest ($2{,}045$ samples) and smallest ($27$ samples) clients.

\textit{Federated learning.} Each client trains locally with SGD or DP-SGD and transmits updates to the Flower FedAvg server, which aggregates via Eq.~\eqref{eq:fedavg} over $R{=}50$ rounds with $E{=}5$ local epochs per round, learning a generalized representation of normal cardiac morphology without accessing raw patient records.

\textit{Differential privacy.} Each client clips per-sample gradients to $\ell_2$ norm $C{=}1.0$ and adds Gaussian noise $\mathcal{N}(0, \sigma^2 C^2 \mathbf{I})$ following Eq.~\eqref{eq:dpsgd}, where $\sigma$ is tuned to achieve a target $\varepsilon$ at $\delta{=}10^{-5}$ over $R=50$ rounds, with accounting via the R\'enyi-DP framework in Opacus.

\textit{Post-training deployment.} The best FP32 checkpoint is quantized to INT8 via dynamic PTQ and benchmarked on Raspberry~Pi~4. Dynamic PTQ requires no retraining or calibration data, making it practical in privacy-sensitive environments where calibration data may not be centrally available.

\vspace{-1mm}
\subsection{Autoencoder Architectures}

Three architectures conform to a shared \texttt{BaseAutoencoder} interface, enabling seamless integration with FedAvg and Opacus. The \emph{VanillaAE} uses fully-connected layers ($12{,}000 \!\to\! 512 \!\to\! 256 \!\to\! 64 \!\to\! 128$ bottleneck). The \emph{ConvAE} uses four 1D convolutional encoder blocks (channels [$32,64,128,256$], kernels [$7,7,5,5$], stride\,$2$, GroupNorm), with a mirrored transposed-convolution decoder. The \emph{VAE} shares the ConvAE encoder but outputs $\boldsymbol{\mu}$ and $\log\boldsymbol{\sigma}^2$, sampling $\mathbf{z} = \boldsymbol{\mu} + \boldsymbol{\sigma} \odot \boldsymbol{\varepsilon}$ via the reparameterization trick~\cite{Var_AE}, with $\mathcal{L} = \text{MSE} + \beta\,D_{\text{KL}}$ ($\beta{=}0.5$, selected for best federated F1/AUPRC balance). All architectures replace BatchNorm with GroupNorm~\cite{Exp_Set5} and use non-inplace activations to satisfy Opacus per-sample gradient requirements.

\vspace{-1mm}
\section{Experimental Setup}
\label{sec:setup}

Our evaluation answers three questions in sequence: does federated learning preserve detection quality under non-IID partitioning; what does formal differential privacy cost in AUROC and per-class behavior; and does the resulting model remain deployable on a constrained edge device. All experiments use a single fixed dataset, a shared evaluation pipeline, and three random seeds across the three autoencoder families.

\textbf{Dataset.} PTB-XL (v1.0.3)~\cite{dataset1}, obtained via PhysioNet~\cite{Exp_Set2}, contains $21{,}799$ ten-second 12-lead ECGs at $100$\,Hz from $18{,}885$ patients annotated under SCP-ECG. Retaining codes with likelihood ${\geq}50\%$ under a one-class framing (NORM\,$=$\,$0$; any non-NORM superclass\,$=$\,$1$) yields $20{,}373$ valid labels ($9{,}038$ normal; $11{,}335$ abnormal), with per-class counts: MI\,$4{,}049$; STTC\,$3{,}360$; CD\,$3{,}431$; HYP\,$1{,}305$.

\textbf{Non-IID partitioning.} The $6{,}294$ normal training samples are distributed across $K{=}10$ clients via Dirichlet ($\alpha{=}0.5$), producing a $75.7\times$ volume disparity between the largest ($2{,}045$ samples) and smallest ($27$ samples) clients, reflecting realistic hospital-size variation. Since training is normal-only, the allocation produces volume skew rather than label heterogeneity.

\textbf{Implementation.} All models use PyTorch~\cite{Exp_Set3}. Training uses Adam (lr\,$=$\,$10^{-3}$, weight decay\,$10^{-5}$, batch\,$64$) with CosineAnnealingWarmRestarts ($T_0{=}20$, $T_\text{mult}{=}2$, $\eta_\text{min}{=}10^{-6}$) for ConvAE/VanillaAE and ReduceLROnPlateau (patience\,$7$, factor\,$0.5$) for VAE. Early stopping monitors validation MSE with patience\,$25$ (ConvAE/VanillaAE) or $15$ (VAE). Federated learning uses $R{=}50$ rounds of $E{=}5$ local epochs with fixed-rate Adam and gradient clipping (max\_norm\,$=$\,$1.0$) throughout. Bottleneck $d{=}128$ was selected via ablation (ConvAE AUROC: $0.653$ at $d{=}8$ to $0.771$ at $d{=}128$); the federated VAE uses $d{=}32$ (architecture default). Seeds: $\{42, 123, 456\}$. The complete implementation is available at an anonymous repository.~\cite{anon_repo}.

\textbf{Evaluation.} AUROC and the area under the precision-recall curve (AUPRC) are the primary threshold-independent metrics. Sensitivity, specificity, and F1 are computed at the $95$\textsuperscript{th}-percentile reconstruction error threshold on validation normal samples, prioritizing specificity ($>$$0.96$) for screening. Per-class evaluation pairs all $1{,}407$ test normals with each abnormal subset (MI\,$536$; STTC\,$457$; HYP\,$189$; CD\,$482$) using the same trained model and threshold, ensuring training is never influenced by subcategory labels. Results are reported as mean\,$\pm$\,std across three seeds, with practical significance assessed by effect-size magnitude.

\vspace{-3mm}
\section{Threat Model}
\label{sec:threat}

Because the privacy guarantee is this system's central claim, we dedicate a separate section to scoping it explicitly.

\noindent\textbf{Privacy unit.} A single ECG recording. DP bounds the influence of any one recording on the released global model (example-level DP within each client's local dataset), not user-level DP across a client's patient population, and not cross-client DP. Reported $\varepsilon$ values are per-client, per-training-run budgets composed over all $R{=}50$ rounds with $\delta{=}10^{-5}$.

\noindent\textbf{Trust assumptions.} The aggregation server is \emph{honest-but-curious}: it follows FedAvg honestly but may inspect any client update it receives. Clients are honest participants (no model poisoning or Byzantine behavior). No secure aggregation is deployed, so per-round client updates are visible to the server in the clear; we therefore claim no amplification from channel secrecy, only from DP noise. Client subsampling amplification is not exploited. Integrating DP with secure aggregation~\cite{DP_SecAgg_ECG2025} is a natural extension discussed as future work.

\noindent\textbf{Adversary capabilities and mitigations.} Table~\ref{tab:threats} enumerates the adversaries considered, the capabilities assumed, and whether each is mitigated by the deployed mechanism.

\begin{table}[!t]
\centering
\caption{Threat matrix. DP: per-sample Gaussian mechanism at each client. \checkmark: mitigated up to the $(\varepsilon,\delta)$ budget. $\star$: out of scope; see discussion.}
\label{tab:threats}
\setlength{\tabcolsep}{3pt}\footnotesize
\begin{tabular}{p{2.1cm}p{2.8cm}cc}
\toprule
\textbf{Adversary} & \textbf{Capability} & \textbf{Mitigation} & \textbf{In scope} \\
\midrule
{\small{Curious server}}     & Observes all per-round updates & DP & \checkmark \\
Curious client     & Observes broadcast global model & DP & \checkmark \\
Network observer   & Intercepts update messages & DP & \checkmark \\
Membership-inference attacker & Queries released model & DP & \checkmark \\
Gradient-inversion attacker & Reconstructs samples from updates & DP & \checkmark \\
Byzantine client   & Sends malicious updates & --- & $\star$ \\
Colluding majority & Reconstructs target from aggregates & --- & $\star$ \\
Query-time attacker & Attacks deployed edge model & --- & $\star$ \\
\bottomrule
\end{tabular}
\end{table}

\textbf{Out of scope.} Malicious protocol deviations (model poisoning, Byzantine clients), colluding-majority reconstruction of a target client's local state, and query-time attacks on the deployed edge model. Empirical auditing via membership-inference attacks complements rather than replaces the formal $(\varepsilon,\delta)$ bound and is left to future work.

\vspace{-2mm}
\section{Results \& Evaluation}
\label{sec:results}

We present results along the three evaluation axes introduced above: federated detection quality under non-IID partitioning, the utility cost of formal differential privacy across a five-point $\varepsilon$ sweep, and a seven-configuration component ablation. Results are mean\,$\pm$\,std over three seeds.

\vspace{-2mm}
\subsection{Computation Efficiency}

\begin{figure}[!t]
\centering
\begin{tikzpicture}
\begin{axis}[
  width=0.98\columnwidth, height=3.8cm,
  xlabel={\small Model Size (MB)},
  ylabel={\small AUROC},
  xmin=2, xmax=10,
  ymin=0.78, ymax=0.8,
  ytick={0.78, 0.79, 0.80},
  yticklabel style={font=\tiny},
  xticklabel style={font=\tiny},
  xlabel style={font=\small},
  ylabel style={font=\small},
  legend pos=south east,
  legend style={font=\tiny, row sep=-2pt},
  grid=major, grid style={gray!20},
  clip=false,
]
\addplot[only marks, mark=*, mark size=2.5pt,
         draw=black!70, fill=blue!55] coordinates {
  (5.71,0.788)(8.30,0.795)};
\addlegendentry{FP32}
\addplot[only marks, mark=square*, mark size=2.5pt,
         draw=black!70, fill=red!55] coordinates {
  (2.76,0.787)(3.87,0.794)};
\addlegendentry{INT8}
\draw[-Stealth,gray!60,thin](axis cs:5.71,0.788)--(axis cs:2.76,0.787);
\draw[-Stealth,gray!60,thin](axis cs:8.30,0.795)--(axis cs:3.87,0.794);
\node[font=\tiny,above] at (axis cs:5.71,0.788){ConvAE\,$\star$};
\node[font=\tiny,above] at (axis cs:8.30,0.795){VAE};
\end{axis}
\end{tikzpicture}
\caption{Detection quality versus model footprint for the two convolutional architectures. Each arrow traces the FP32\,$\to$\,INT8 transition for one model; the near-horizontal arrows show that INT8 quantization roughly halves model size while changing AUROC by less than $0.12\%$. The starred ConvAE\,INT8 point ($2.76$\,MB, AUROC\,$0.787$) is the recommended edge configuration, trading a marginal AUROC reduction for the smallest footprint and lowest Pi~4 latency.}
\label{fig:comp_eff}
\end{figure}
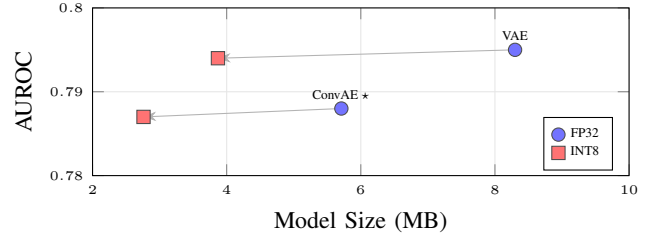

Table~\ref{tab:computation} summarizes storage, Raspberry~Pi~4 latency, and detection quality for FP32 and INT8 models. INT8~PTQ reduces size by $51.7$--$74.9\%$: VanillaAE drops most ($74.9\%$), while ConvAE ($51.7\%$) and VAE ($53.3\%$) retain a proportionally larger convolutional footprint. Pi~4 inference improves by $1.20$--$1.80\times$ with AUROC degradation below $0.12\%$ in every case (Fig.~\ref{fig:comp_eff}).

\begin{table}[!t]
\caption{Computation efficiency across deployment environments.}
\label{tab:computation}
\centering
\renewcommand{\arraystretch}{1.05}
\setlength{\tabcolsep}{3pt}
\footnotesize
\begin{tabular}{llcccc}
\toprule
\textbf{Model} & \textbf{Prec.} &
\textbf{Size} & \textbf{FLOPs} & \textbf{Pi4 Lat.} & \textbf{AUROC} \\
& & \textbf{(MB)} & \textbf{(M)} & \textbf{(ms)} & \\
\midrule
VanillaAE & FP32 & $48.07$ & $25.2$ & $45.2\pm0.1$ & $0.637$ \\
VanillaAE & INT8 & $12.07$ & $25.2$ & $25.2\pm0.1$ & $0.637$ \\
\midrule
ConvAE    & FP32 &  $5.71$ & $85.0$ & $16.9\pm0.1$ & $0.788$ \\
ConvAE    & INT8 &  $2.76$ & $85.0$ & $14.1\pm0.1$ & $0.787$ \\
\midrule
VAE        & FP32 &  $8.30$ & $110.2$ & $21.0\pm0.2$ & $0.795$ \\
VAE        & INT8 &  $3.87$ & $110.2$ & $16.6\pm0.0$ & $0.794$ \\
\bottomrule
\multicolumn{6}{l}{\scriptsize FLOPs unchanged by PTQ. Latency = mean\,$\pm$\,std over 3 seeds.}
\end{tabular}
\end{table}

ConvAE\,INT8 attains AUROC\,$=$\,$0.787$ at $2.76$\,MB and $14.1$\,ms (${\approx}56$\,mJ/inference\footnote{Energy: $E{=}P{\times}t_\text{inf}$, $P{\approx}4.0$\,W sustained CPU load on Pi~4. Direct wattmeter validation remains future work.}). VAE achieves marginally higher AUROC ($0.794$) at $30\%$ greater FLOPs. Pi~4 benefits more from INT8 than the x86-64 reference PC ($1.20$--$1.80\times$ vs.\ $0.69$--$1.46\times$), confirming that reduced precision has a greater practical effect under tight edge constraints~\cite{Comput_Result}. Per-inference energy for INT8 variants is approximately $100.6$\,mJ (VanillaAE), $56.4$\,mJ (ConvAE), and $66.6$\,mJ (VAE), with savings of $16.6$--$44.3\%$ over FP32.

\vspace{-1mm}
\subsection{Privacy and Utility Trade-off}
\label{sec:privacy_results}

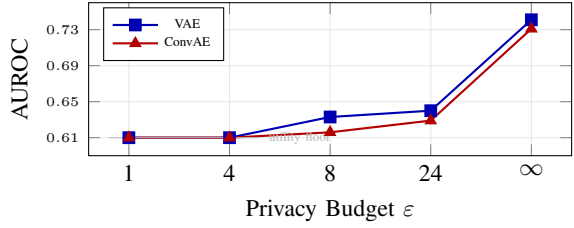
\begin{figure}[!t]
\centering
\begin{tikzpicture}
\begin{axis}[
  width=0.90\columnwidth, height=3.6cm,
  xlabel={\small Privacy Budget $\varepsilon$},
  ylabel={\small AUROC},
  xtick={1,2,3,4,5},
  xticklabels={1,4,8,24,$\infty$},
  xticklabel style={font=\small},
  yticklabel style={font=\tiny},
  xlabel style={font=\small},
  ylabel style={font=\small},
  ymin=0.590, ymax=0.760,
  ytick={0.61,0.65,0.69,0.73},
  legend pos=north west,
  legend style={font=\tiny, row sep=-2pt},
  grid=major, grid style={gray!18},
]
\addplot[blue!70!black, mark=square*, thick, mark size=2pt]
  coordinates{(1,0.610)(2,0.610)(3,0.633)(4,0.640)(5,0.741)};
\addlegendentry{VAE}
\addplot[red!70!black, mark=triangle*, thick, mark size=2pt]
  coordinates{(1,0.610)(2,0.610)(3,0.616)(4,0.629)(5,0.731)};
\addlegendentry{ConvAE}
\draw[gray!50, thin] (axis cs:0.8,0.610) -- (axis cs:2.3,0.610)
  node[right, font=\tiny, text=gray!55]{utility floor};
\end{axis}
\end{tikzpicture}
\caption{Privacy and utility trade-off. The elbow between $\varepsilon{=}4$ and $\varepsilon{=}8$ marks the transition into the utility floor; see Section~\ref{sec:discussion} for our recommended operating point.}
\label{fig:privacy_utility}
\end{figure}

Table~\ref{tab:dp_results} reports metrics under DP-SGD for both top-performing architectures. Without DP, VAE and ConvAE reach AUROC\,$=$\,$0.741$ and $0.731$ respectively. Stronger privacy monotonically degrades performance, with both converging to a utility floor (AUROC\,${\approx}\,0.610$) at $\varepsilon{=}1$ where gradient noise overwhelms reconstruction learning. Between $\varepsilon{=}4$ and $\varepsilon{=}8$, VAE AUROC increases by only $0.022$ while the formal guarantee doubles (Fig.~\ref{fig:privacy_utility}). DP-SGD adds only $3$--$6$\,s/round over the non-private baseline, confirming that the privacy penalty is one of utility, not computation.

\begin{table}[!t]
\centering
\caption{Privacy and utility trade-off under DP-SGD. $\varepsilon{=}\infty$: no differential privacy.}
\label{tab:dp_results}
\renewcommand{\arraystretch}{0.95}
\setlength{\tabcolsep}{3pt}
\footnotesize
\begin{tabular}{llccccc}
\toprule
\textbf{Model} & $\boldsymbol{\varepsilon}$ &
\textbf{AUC} & \textbf{AUPRC} &
\textbf{Sens.} & \textbf{Spec.} & \textbf{s/rnd} \\
\midrule
VAE & $\infty$ & $0.741$ & $0.769$ & $0.086$ & $0.992$ &  $8.5$ \\
VAE & $1$      & $0.610$ & $0.659$ & $0.073$ & $0.976$ & $11.5$ \\
VAE & $4$      & $0.610$ & $0.659$ & $0.073$ & $0.976$ & $13.5$ \\
VAE & $8$      & $0.633$ & $0.679$ & $0.074$ & $0.978$ & $14.2$ \\
VAE & $24$     & $0.640$ & $0.687$ & $0.076$ & $0.979$ & $13.4$ \\
\midrule
ConvAE & $\infty$ & $0.731$ & $0.762$ & $0.085$ & $0.991$ & $10.9$ \\
ConvAE & $1$      & $0.610$ & $0.659$ & $0.073$ & $0.976$ & $12.9$ \\
ConvAE & $4$      & $0.610$ & $0.659$ & $0.073$ & $0.976$ & $13.5$ \\
ConvAE & $8$      & $0.616$ & $0.665$ & $0.074$ & $0.978$ & $14.3$ \\
ConvAE & $24$     & $0.629$ & $0.677$ & $0.074$ & $0.978$ & $14.3$ \\
\bottomrule
\multicolumn{7}{l}{\scriptsize Higher $\varepsilon$ = weaker privacy; lower $\varepsilon$ = stronger privacy.}
\end{tabular}
\end{table}

\noindent\textbf{Per-class impact of differential privacy.} The centralized VAE exhibits posterior collapse at all tested $\beta$ values, where the KL term suppresses latent code utilization and the decoder learns to ignore $\mathbf{z}$. Under federation, heterogeneous client updates from the non-IID Dirichlet partitioning act as implicit regularization, preventing this collapsed mode; we return to this in Section~\ref{sec:discussion}.

Per-class AUROC values ($0.45$--$0.49$) cluster near the random-classifier baseline of $0.50$, as expected for unsupervised one-class detectors distinguishing anomalies through aggregate reconstruction error rather than class-discriminative features. DP does not degrade any diagnostic category catastrophically. HYP remains the hardest class (AUROC $0.447$--$0.454$) due to voltage-amplitude changes that overlap substantially with normal variation rather than privacy noise~\cite{Exp_Set7}. CD shows a slight regularization benefit under DP (AUROC\,$0.496$ at $\varepsilon{=}24$ vs.\ $0.482$ at $\varepsilon{=}\infty$); MI and STTC remain stable (range ${<}\,0.005$). The aggregate AUROC of $0.74$--$0.78$ confirms meaningful separation overall; per-class figures identify which conditions would benefit most from future refinement.

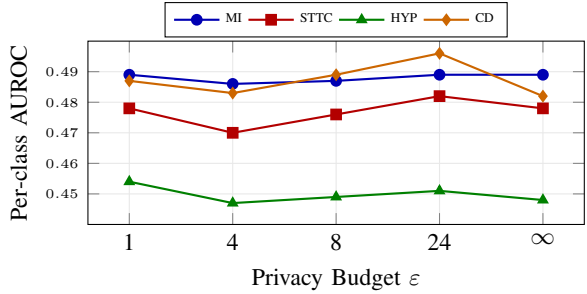
\begin{figure}[!t]
\centering
\begin{tikzpicture}
\begin{axis}[
  width=0.92\columnwidth, height=4.0cm,
  xlabel={\small Privacy Budget $\varepsilon$},
  ylabel={\small Per-class AUROC},
  xtick={1,2,3,4,5},
  xticklabels={1,4,8,24,$\infty$},
  xticklabel style={font=\small},
  yticklabel style={font=\tiny},
  xlabel style={font=\small},
  ylabel style={font=\small},
  ymin=0.440, ymax=0.500,
  ytick={0.45,0.46,0.47,0.48,0.49},
  legend pos=outer north east,
  legend style={font=\tiny, row sep=-2pt, legend columns=4,
                at={(0.5,1.02)}, anchor=south},
  grid=major, grid style={gray!18},
]
\addplot[blue!70!black, mark=*, thick, mark size=1.8pt]
  coordinates{(1,0.489)(2,0.486)(3,0.487)(4,0.489)(5,0.489)};
\addlegendentry{MI}
\addplot[red!70!black, mark=square*, thick, mark size=1.8pt]
  coordinates{(1,0.478)(2,0.470)(3,0.476)(4,0.482)(5,0.478)};
\addlegendentry{STTC}
\addplot[green!50!black, mark=triangle*, thick, mark size=1.8pt]
  coordinates{(1,0.454)(2,0.447)(3,0.449)(4,0.451)(5,0.448)};
\addlegendentry{HYP}
\addplot[orange!80!black, mark=diamond*, thick, mark size=1.8pt]
  coordinates{(1,0.487)(2,0.483)(3,0.489)(4,0.496)(5,0.482)};
\addlegendentry{CD}
\end{axis}
\end{tikzpicture}
\caption{Per-class AUROC across diagnostic categories under
varying privacy budgets. HYP is consistently the hardest class;
CD shows a slight regularization benefit under DP. No category
suffers catastrophic degradation.}
\label{fig:perclass_dp}
\end{figure}

Fig.~\ref{fig:perclass_dp} visualizes these trends. No diagnostic class suffers catastrophic degradation, supporting the viability of DP-SGD for population-level screening.

\vspace{-1mm}
\subsection{Architecture Comparison and Ablation}

Table~\ref{tab:ae_comparison} compares all three architectures. ConvAE leads in centralized AUROC ($0.764$), as 1D convolutions capture ECG temporal morphology more effectively than fully-connected layers (VanillaAE, $0.650$) or the VAE ($0.596$, impaired by posterior collapse). Under federation the VAE registers the largest gain ($+0.165$, to $0.761$): non-IID heterogeneity acts as implicit regularization on the stochastic latent space, rescuing the collapsed baseline. ConvAE improves modestly ($+0.018$ to $0.782$); VanillaAE is unchanged ($0.648$), confirming that fully-connected architectures lack the inductive bias to benefit from data heterogeneity.

\begin{table}[!t]
\caption{Architecture comparison: local vs.\ federated (mean\,$\pm$\,std, three seeds).}
\label{tab:ae_comparison}
\centering
\renewcommand{\arraystretch}{1.0}
\setlength{\tabcolsep}{2.5pt}
\footnotesize
\begin{tabular}{llccccc}
\toprule
\textbf{Model} & \textbf{Setting} &
\textbf{AUC} & \textbf{AUPRC} &
\textbf{Sens.} & \textbf{Spec.} & \textbf{F1} \\
\midrule
VanillaAE & Local & $0.650_{\pm 0.000}$ & $0.721$ & $0.218$ & $0.963$ & $0.349$ \\
ConvAE    & Local & $0.764_{\pm 0.029}$ & $0.790$ & $0.246$ & $0.967$ & $0.386$ \\
VAE        & Local & $0.596_{\pm 0.007}$ & $0.651$ & $0.110$ & $0.959$ & $0.192$ \\
\midrule
VanillaAE & Fed.\ & $0.648_{\pm 0.000}$ & $0.718$ & $0.204$ & $0.964$ & $0.331$ \\
ConvAE    & Fed.\ & $0.782_{\pm 0.004}$ & $0.801$ & $0.245$ & $0.969$ & $0.385$ \\
VAE        & Fed.\ & $0.761_{\pm 0.005}$ & $0.784$ & $0.245$ & $0.961$ & $0.383$ \\
\bottomrule
\end{tabular}
\vspace{-2mm}
\end{table}

\noindent\textbf{Component ablation.} Table~\ref{tab:ablation} presents the seven-configuration ablation on the VAE. At matched $d{=}32$, federation improves VAE AUROC from $0.596$ to $0.743$ ($+0.147$), with non-IID volume skew ($75.7\times$ across $K{=}10$ clients) preventing posterior collapse. \textit{Privacy} (Cfg.\,$2\!\to\!3$): DP-SGD at $\varepsilon{=}4$ costs $-0.124$ AUROC; tightening to $\varepsilon{=}1$ (Cfg.\,$6$) adds only $-0.009$ further, confirming the utility floor. \textit{Quantization} (Cfg.\,$2\!\to\!4$): INT8 reduces size by $53.4\%$ and Pi~4 latency by $20.8\%$ with negligible AUROC change ($+0.006$). \textit{Combined} (Cfg.\,$3\!\to\!5$): adding INT8 on top of DP-SGD incurs at most $-0.003$ additional AUROC, confirming that the two penalties are independent and do not compound, making the combined configuration strictly preferable for edge deployment.

\begin{table}[!t]
\centering
\caption{Component ablation (VAE; mean\,$\pm$\,std, seeds \{$42,123$\}). Cfg.\,$1$ = centralized FP32 baseline.}
\label{tab:ablation}
\renewcommand{\arraystretch}{1.05}
\setlength{\tabcolsep}{2.5pt}
\footnotesize
\begin{tabular}{ccccccc}
\toprule
\textbf{Cfg.} & \textbf{FL} & \textbf{INT8} &
\textbf{DP\,($\varepsilon$)} &
\textbf{AUROC} & \textbf{Size} & \textbf{Pi4\,Lat.} \\
\midrule
$1$ & \ding{55} & \ding{55} & $\infty$  & $0.747_{\pm 0.005}$ & $8.30$ & $21.0$ \\
$2$ & \ding{51} & \ding{55} & $\infty$  & $0.743_{\pm 0.001}$ & $8.30$ & $21.0$ \\
$3$ & \ding{51} & \ding{55} & $4$       & $0.619_{\pm 0.008}$ & $8.30$ & $21.0$ \\
$4$ & \ding{51} & \ding{51} & $\infty$  & $0.749_{\pm 0.013}$ & $3.87$ & $16.6$ \\
$5$ & \ding{51} & \ding{51} & $4$       & $0.616_{\pm 0.005}$ & $3.87$ & $16.6$ \\
$6$ & \ding{51} & \ding{55} & $1$       & $0.610_{\pm 0.000}$ & $8.30$ & $21.0$ \\
$7$ & \ding{51} & \ding{51} & $1$       & $0.611_{\pm 0.001}$ & $3.87$ & $16.6$ \\
\bottomrule
\multicolumn{7}{l}{\scriptsize Size in MB; latency in ms on Raspberry~Pi~4. Cfg.\,$1$: $d{=}128$; fed.: $d{=}32$.}
\end{tabular}
\vspace{-2mm}
\end{table}

\section{Discussion}
\label{sec:discussion}

We discuss the recommended DP operating point, the utility floor, federation as implicit regularization, the independence of DP and quantization penalties, deployment implications, and limitations.

\noindent\textbf{Recommended operating point ($\varepsilon{=}4$).} We recommend $\varepsilon{=}4$ on four grounds. \textit{(i) Utility floor:} reducing $\varepsilon$ below $4$ yields no measurable AUROC gain. \textit{(ii) Marginal cost:} halving $\varepsilon$ from $8$ to $4$ costs only $-0.022$ AUROC while doubling the formal guarantee. \textit{(iii) Clinical-data strictness:} non-medical deployments operate at $\varepsilon\in[8,16]$ or higher~\cite{Ponomareva2023}, but ECG morphology is quasi-biometric and re-identifiable~\cite{Deanon_ECG2025}, and adversarial linkage to auxiliary records amplifies real-world leakage beyond the nominal $e^{\varepsilon}$ ratio. \textit{(iv) Community alignment:} $\varepsilon{=}4$ matches comparable healthcare FL studies~\cite{Paper5,Kaissis2020}. Under the honest-but-curious model of Section~\ref{sec:threat}, this budget bounds the likelihood ratio at $e^{4}{\approx}54.6$, consistent with HIPAA Safe Harbor de-identification guidance for quasi-identifiers.

\noindent\textbf{The utility floor phenomenon.} The plateau at AUROC\,${\approx}\,0.610$ for $\varepsilon \leq 4$ (Table~\ref{tab:dp_results}) is a practically important finding. Below this threshold, DP noise exceeds the reconstruction-error gradient signal, causing the autoencoder to converge to a trivially smooth reconstruction mapping all inputs to approximately the population mean. This floor is specific to reconstruction-based detectors; supervised FL pipelines show more gradual degradation because cross-entropy maintains stronger per-class gradients under noise~\cite{Paper5}. Adaptive noise scheduling may lift this floor while preserving formal cumulative $\varepsilon$ guarantees.

\noindent\textbf{Federation as implicit regularization.} The VAE improvement under federation ($+0.165$ AUROC) stems from the centralized VAE suffering posterior collapse. Heterogeneous client updates from non-IID Dirichlet partitioning prevent the global model from settling into this collapsed mode, suggesting regularization benefits of FL beyond its primary privacy motivation.

\noindent\textbf{Independence of DP and quantization penalties.} DP and INT8 quantization penalties are empirically independent (Table~\ref{tab:ablation}, Cfg.\,$3$ vs.\ $5$). System designers can enable formal privacy guarantees and aggressive compression simultaneously without fearing multiplicative degradation, most likely because DP-SGD perturbs training while PTQ compresses already-trained weights, acting on disjoint parts of the pipeline.

\noindent\textbf{Deployment considerations.} The recommended configuration (ConvAE\,INT8: AUROC\,$=$\,$0.787$, $2.76$\,MB, $14.1$\,ms, ${\approx}56$\,mJ) supports real-time screening at ${\approx}71$ inferences/second on Raspberry~Pi~4, enabling one classification every 10\,s with ${<}1\%$ CPU utilization under continuous 100\,Hz monitoring, leaving headroom for signal acquisition and wireless transmission.

\noindent\textbf{Client scalability.} Preliminary experiments with $K \in \{5, 10, 20\}$ (co-varying $\alpha$ as $0.1, 0.5, 1.0$) show AUROC degrades by at most $0.025$ from $K{=}5$ to $K{=}20$, while per-round time drops at $K{=}20$ ($6.9$--$7.2$\,s vs.\ $12.3$--$16.7$\,s). Disentangling $K$ and $\alpha$ is left as future work.

\vspace{-1mm}
\subsection{Ethical Considerations}
\label{sec:ethics}
Deploying AI within clinical workflows raises five ethical issues directly informed by our results. \textit{(i) Privacy is non-negotiable for ECG.} ECG waveforms are quasi-biometric, and linkage attacks have been demonstrated in practice~\cite{Deanon_ECG2025,ECG_Attacks2025}. Federation without formal privacy leaves a realistic attack surface via gradient inversion and membership inference, making the $(\varepsilon,\delta)$-guarantee a prerequisite rather than an enhancement. \textit{(ii) Federation is especially valuable in healthcare.} Cardiac morphology varies systematically across demographics and protocols; FL permits diverse participation without raw-data transfer, and the $75.7\times$ non-IID volume skew we observed reflects real hospital-network properties. \textit{(iii) Population screening, not condition-specific diagnosis.} Per-class AUROC values clustering near $0.50$ follow directly from the unsupervised one-class objective. The system is suited to screening and triage for cardiologist review, and must not be presented as a condition-specific diagnostic tool. \textit{(iv) Tolerable error margins.} Our $95$\textsuperscript{th}-percentile operating point keeps false positives below $4\%$ on normal recordings, but real deployments must calibrate the threshold to the receiving pathway's tolerance. \textit{(v) Honest aggregator is insufficient for production.} Our formal claim assumes an honest-but-curious server; extending the system under stronger threat models~\cite{DP_SecAgg_ECG2025} is a precondition for responsible clinical use.

\vspace{-1mm}
\subsection{Limitations}

Experiments are restricted to a single dataset (PTB-XL); generalizability to corpora such as MIT-BIH or fewer-lead wearable form factors has not been validated. Federated heterogeneity is simulated via Dirichlet partitioning rather than observed from real deployments. Energy consumption is estimated via a latency-proportional proxy rather than direct wattmeter measurement, and peak runtime memory was not directly profiled on Pi~4. The component ablation uses two seeds rather than three for computational reasons. With $n{=}3$ seeds per configuration the sample size is insufficient for non-parametric significance tests (minimum achievable Wilcoxon $p{=}0.25$), so we rely on effect-size magnitude. No systematic threshold sensitivity analysis was performed; however, AUROC and AUPRC are threshold-independent, so the reported detection capability is decoupled from any single operating point. Principled threshold calibration to the target pathway's false-positive tolerance is a priority for follow-up work.

\section{Conclusion and Future Work}
\label{sec:conclusion}

Continuous cardiac monitoring can only save lives if the underlying pipeline respects patient privacy and fits on the hardware patients actually carry. We built and measured an end-to-end federated system that trains unsupervised autoencoders across ten simulated hospitals on real 12-lead ECG data, enforces example-level differential privacy with a clearly stated threat model, compresses the resulting model to INT8, and runs it on a Raspberry~Pi~4 at real-time latencies.

The three mechanisms compose gracefully. Federated learning preserves or improves detection quality over the centralized baseline; a budget of $\varepsilon{=}4$ offers a meaningful formal guarantee at a tolerable utility cost; and INT8 quantization halves model size without measurably harming accuracy. Crucially, the DP and quantization penalties are empirically independent, so practitioners need not choose between a strong privacy claim and a small edge footprint. The system is deliberately scoped as a population-level screening and triage tool rather than a standalone diagnostic.

Six directions for future work follow from the results. \emph{Adaptive noise scheduling} may lift the utility floor observed below $\varepsilon{=}4$. \emph{Cross-dataset validation} on corpora such as MIT-BIH would strengthen generalizability. \emph{Empirical privacy auditing} via membership-inference attacks would complement the formal guarantee. \emph{Hybrid supervised-and-unsupervised objectives} may improve per-class detection for clinically critical conditions without abandoning the label-free paradigm. \emph{Secure aggregation.} The honest-but-curious server assumption of Section~\ref{sec:threat} leaves per-round client updates visible to the aggregator, which our $(\varepsilon,\delta)$-DP guarantee mitigates but does not eliminate. Integrating the SecAgg+ protocol~\cite{Bell2020SecAggPlus}, available as a workflow in the Flower framework we already use, would ensure the server observes only the aggregated update and tolerate client dropout via $t$-of-$n$ secret sharing. Quantifying the resulting bandwidth and round-time overhead of mask generation on Raspberry~Pi~4 class hardware, together with explainable components~\cite{XAI_FL_Edge2025} and homomorphic encryption~\cite{HE_DP_FL_ECG2026}, would strengthen the deployment claim of this work. Lastly, \emph{Distributed Differential Privacy} (DDP) \cite{Dwork_2006_Our} which addresses the preceding directions assumption of a single trusted aggregator that holds the composed privacy budget across rounds, an assumption the broader literature increasingly challenges. As presented in \cite{fu_2025_differentially}, replacing centralised Gaussian perturbation with per-client discrete noise mechanisms such as the Skellam or Discrete Gaussian distribution, compatible with modular-arithmetic aggregation protocols, allows the server to observe only encrypted aggregates whose summed noise satisfies $(\varepsilon, \delta)$-DP. The privacy amplification inherent to the shuffle model, yielding a centralised loss $\varepsilon_c \ll \varepsilon_l$, could further permit operation below the $\varepsilon{=}4$ utility floor identified above without sacrificing detection performance. Characterising this trade-off on the constrained hardware targeted by this work remains an open and practically significant direction.

{
\linespread{0.85}
\printbibliography
}

\end{document}